\documentstyle[sprocl]{article}

\def\Journal#1#2#3#4{{#1} {\bf #2}, #3 (#4)}

\def\NPB{{\em Nucl. Phys.} B}
\def\NPA{{\em Nucl. Phys.} A}
\def\PLB{{\em Phys. Lett.}  B}
\def\PRL{\em Phys. Rev. Lett.}
\def\PRD{{\em Phys. Rev.} D}

\def\be{\begin{equation}}
\def\ee{\end{equation}}
\def\bea{\begin{eqnarray}}
\def\eea{\end{eqnarray}}

\def\be{\begin{equation}}
\def\ee{\end{equation}}
\def\bea{\begin{eqnarray}}
\def\eea{\end{eqnarray}}

\begin{document}

\title{SUM RULES FOR HIGHER-TWIST PARTON DISTRIBUTIONS}

\author{M. Burkardt}

\address{New Mexico State University, 
Las Cruces,\\ NM 88003, USA\\E-mail: burkardt@nmsu.edu} 


\maketitle\abstracts{In deep-inelastic scattering experiments,
there is a general connection between 
subtractions in dispersion relations, violations of sum-rules and 
$\delta$-functions in parton distribution functions.
It is explained why one might expect a small violation
in sum rules for the twist-3 distribution functions $g_T(x)$ and $h_L(x)$
when the sum-rules are applied to $x\neq 0$ data only.
The non-perturbative predictions are studied in the context of a one-loop 
model.
}

\section{Introduction}
In the theoretical analysis of deep-inelastic scattering (DIS), one usually 
applies the operator product
expansion (OPE) for $Q^2 \rightarrow \infty$ 
to the Compton amplitude --- a procedure which
implicitly involves analytic continuation of the Compton amplitude 
to the regime where ${Q^2}>{2M\nu}$. Formally this step is accomplished by 
invoking dispersion relations. However, at least in principle,
it may happen that there appear subtractions in these dispersion relations
which then manifest themselves as a violation of naive\footnote{Naive, because
they are derived assuming that there are no subtractions.}
sum-rules.

In this note, we study the issue of subtractions in the context of the 
higher-twist parton distributions $h_L(x)$ and $g_T(x)$, which are defined
as correlation functions in a light-like direction \cite{ji}
\bea
S\cdot n\, h_L(x) &\equiv& \frac{1}{2M} \int \frac{d \lambda}{2\pi}
e^{i\lambda x}
\langle PS |\bar{\psi}(0)\sigma^{+-} i \gamma_5 \psi (\lambda n)|PS \rangle
\nonumber\\
S_\perp\, g_T(x) &\equiv& \frac{1}{2p^+} \int \frac{d \lambda}{2\pi}
e^{i\lambda x}
\langle PS |\bar{\psi}(0)\gamma_\perp\gamma_5 \psi (\lambda n)|PS \rangle,
\label{eq:def}
\eea
where $p^\pm=\frac{1}{\sqrt{2}}\left(p^0\pm p^3\right)$ denotes the
usual light-cone coordinates and the light-like vector $n^\mu$ projects out
$n \cdot A = A^+$ for all 4-vectors $A^\mu$.

Like the more familiar polarized twist-two distributions $g_1(x)$ and 
$h_1(x)$, these twist-three
distributions are important physical quantities which summarize low energy
properties of the nucleon in high-energy scattering processes \cite{ji}:
$g_T(x)$ appears as a $\frac{1}{Q}$ correction in DIS \cite{g2exp}, while 
$h_L(x)$ can be measured for example 
in the nucleon-nucleon polarized Drell-Yan process.

Upon integrating over $x$ in Eq. (\ref{eq:def}), comparing with similar
definitions for $g_1$ and $h_1$, and using Lorentz invariance 
one can thus derive the sum-rules \cite{ji}\cite{me} 
\bea
\int_{-1}^1 dx h_L(x) &=& \int_{-1}^1 dx h_1(x) \label{eq:hsr}\\
\int_{-1}^1 dx g_T(x) &=& \int_{-1}^1 dx g_1(x) \label{eq:gsr}.
\eea
Eq. (\ref{eq:gsr}) is also known as the Burkhardt-Cottingham (BC) sum-rule 
\cite{bc}. For the parton distribution functions, defined as light-cone 
correlations (\ref{eq:def}), these sum-rules (\ref{eq:hsr},\ref{eq:gsr})
are a direct consequence of Lorentz invariance, and therefore hardly
anybody would question their validity (assuming the integrals converge).
However, the issue here is whether these sum rules are also valid when
applied to experimental data! Using the operator product expansion in
the Bjorken limit, one can show that for $x \neq 0$, the parton distribution 
functions (defined as a light-cone correlations) 
agree with the corresponding experimentally measured structure 
functions. In DIS experiments, the point $x=0$ must always be excluded since
it corresponds to $Q^2=0$. In practice, DIS experiments
can only measure
\be
\lim_{\varepsilon \rightarrow 0}\int_\varepsilon^1 dx \left[g_T(x)
+ g_T(-x)\right]
\ee
and similarly for the other distributions.
Therefore, what really is being tested when one tests the above sum-rules
is neither Lorentz invariance nor the OPE but whether or not
Eqs. (\ref{eq:hsr},\ref{eq:gsr}) receive a nonzero contribution from
the point $x=0$, i.e. whether or not $h_L(x)$ and  $g_T(x)$ 
[{\em defined} as in Eq. (\ref{eq:def})] contain
$\delta(x)$-type singularities at the origin.

The paper is organized as follows.
In the next section, we will explain the general connections between 
subtractions in dispersion relations, violations of sum-rules and 
$\delta$-functions in parton distribution functions.
The main question is whether a situation as described above does actually
occur or whether it is only of academic interest.
In order to understand the implications for QCD, we will 
use the moment relations derived from the OPE to derive relations between
the small $x$ behaviour of polarized twist-2 distributions 
($g_1(x)$ and $h_1(x)$) and the coefficient of the $\delta$-function of
polarized twist-3 distributions [$h_L(x)$ and $g_\perp(x)$]. 
Finally, these predictions are illustrated using a one-loop 
model.

\section{Real and Imaginary parts of the Compton Amplitude in DIS}
In this section, general connections between subtraction constants, violation
of sum-rules and $\delta(x)$-terms in parton distributions will be discussed.
For this purpose, we will denote $G(x,Q^2)$ and $T(x,Q^2)$ the
real and imaginary part respectively of some generic forward Compton amplitude,
where $x=\frac{Q^2}{2M\nu}$.
The Compton amplitude is an analytic function of $x$, except
along a cut from $0<x<1$ (the cut appears along the physical
region for DIS!). 
Therefore, it should be possible to relate the
real and imaginary part using a dispersion relation
\be
T(x^\prime,Q^2) = \frac{1}{\pi}\int_0^1 dx \frac{x^\prime}{{x^\prime}^2
-x^2} G(x,Q^2) + p\left(\frac{1}{x^\prime},Q^2\right),
\label{eq:disp}
\ee
where
$p(z,Q^2)$ is a polynomial in $z$ and where $|x^\prime|>1$.
In the language of dispersion theory, the `polynomial subtraction'
is related to `J=0 fixed poles'.

In the theoretical analysis of DIS one usually applies the operator 
product expansion (OPE) for $|x^\prime|>1$, yielding
\be
T(x^\prime,Q^2) = \sum_{n=0,2,..} \frac{1}{{x^\prime}^{n+1}} a_n
\ee
where the $a_n$ can be expressed as matrix elements of the form
\be
a_n = \langle P,S|\bar{\psi} \Gamma D^{n-1} \psi |P,S \rangle ,
\label{eq:an}
\ee
which appear in the OPE of $T J_\mu(\xi) J_\nu (0)$.
Here, $\Gamma$ is some Dirac matrix which depends on the particular
structure function (e.g. polarized or unpolarized DIS) and
$D^n$ denotes an $n-th$ order covariant derivative. In order to keep the 
discussion as general as possible, $\Gamma$ will not be specified here any 
further.
 
In the most simple case, i.e.
when there is no subtraction in Eq. (\ref{eq:disp}) and $p(z)\equiv0$, 
the $a_n$ can be expressed through the moments of $G$ by 
expanding the geometric series in Eq. (\ref{eq:disp}), yielding
\be
a_n = \frac{1}{\pi} \int_0^1 dx x^{n-1} G(x,Q^2).
\label{eq:moment0}
\ee
However, in the following we want to investigate what happens
if $p(z)\neq 0$. As a specific example, let us assume that
$p\left(\frac{1}{x^\prime}\right)= \frac{c}{x^\prime}$, where
$c$ is some (possibly $Q^2$ dependent) constant.
Such a scenario has several important consequences:
\begin{itemize}
\item First of all, the simple relation between the $a_n$ and
$\int_0^1 dx x^{n-1} G(x,Q^2)$ (\ref{eq:moment0}) is spoiled
for the lowest moment and instead one finds
\be
a_n = \left\{
\begin{array}{cl}
\frac{1}{\pi} \int_0^1 dx x^{n-1} G(x,Q^2)
& \quad \quad n=3,5,..\\[2.ex]
\frac{1}{\pi} \int_0^1 dx x^{n-1} G(x,Q^2)\quad + \quad c
& \quad \quad n=1
\end{array}\right.
\ee
\item As a consequence, the naive sum-rule $a_1=\frac{1}{\pi} 
\int_0^1 dx G(x,Q^2)$ is of course violated if $c\neq 0$
\item Suppose that one defines a parton distribution function
$g(x)$ through the moments, i.e. by requiring that
\be
a_n \stackrel{!}{=} \frac{1}{\pi} {\cal M}_n\left[g\right]
\equiv \frac{1}{\pi} \int_{-1}^1 dx x^{n-1}
g(x,Q^2) \quad \quad n\geq 0
\ee
then $g(x,Q^2)$ differs from the experimentally measured $G(x,Q^2)$
by a $\delta$ function at the origin, i.e.
\be
g(x,Q^2)+g(-x,Q^2) = G(x,Q^2)+c\pi\delta(x).
\ee
This result should be intuitively clear since the above subtraction 
affects only the lowest moment, which means that there must be a
$\delta$-function present.
\end{itemize}
To summarize the above discussion, we emphasize that there is no problem with
the OPE if a sum-rule derived for parton distributions fails when
applied to experimental data. All it means is that the corresponding
dispersion relation has a subtraction and the parton distribution function
(defined as a generalized function through the light-cone moments) 
has a $\delta$-function at the origin, which is not present in the experimental
DIS data. Therefore, an experiment which measures
$\lim_{\varepsilon \rightarrow 0} \int_\varepsilon^1 dx G(x,Q^2)$ would miss
the $\delta$ function and hence a ``violation'' of the sum-rule would
be observed.

There exist a number of toy models where such $\delta(x)$ terms
have been observed in parton distribution for models in
2 and 4 dimensions. Because of lack of space, the reader is referred to these
references \cite{toy} and we focus in the following on the most important
question, namely whether such $\delta$ functions occur in $QCD_{3+1}$.

\section{OPE, Moments and $\delta$-Functions in QCD}
In this section we start from the relation among the moments of
$h_L(x)$, $h_1(x)$ and $g_1(x)$ \cite{ji}
(note that ${\cal M}_n[\tilde{h}_L]=0$ for $n \leq 2$).

\be
{\cal M}_n\left[h_L\right] = 
\left\{\begin{array}{ll} {\cal M}_n\left[h_1\right] &
\quad n = 1 
\\[2.ex]
\frac{2}{n+1}{\cal M}_n\left[h_1\right] + 
{\cal M}_n\left[\tilde{h}_L\right] + \frac{m_q}{M} \frac{n-1}{n+1}
{\cal M}_{n-1}\left[g_1\right] \quad & \quad n \geq 2 \end{array}\right.,
\ee
where $\tilde{h}_L$ is the interaction-dependent twist-three part of $h_L$.
Upon inverting the moment relation, one finds \cite{you}
\bea
h_L(x,\mu^2) &=& 2x\!\!\int_x^1\!\!\!dy \frac{h_1 (y,\mu^2)}{y^2}
+ \tilde{h}_L(x,\mu^2) 
+ \frac{m_q}{M} \left[ \frac{g_1(x,\mu^2)}{x} -2x \!\!\int_x^1 \!\!\!dy
  \frac{g_1(y,\mu^2)}{y^3}\right] \nonumber\\ & &\label{eq:ope1a}\\
h_L(x,\mu^2) &=& -2x\!\!\int_{-1}^x\!\!\!\!\!dy \frac{h_1 (y,\mu^2)}{y^2}
+ \tilde{h}_L(x,\mu^2) 
+ \frac{m_q}{M} \!\left[ \frac{g_1(x,\mu^2)}{x} +2x \!\!\!\int_{-1}^x 
\!\!\!\!\!dy
  \frac{g_1(y,\mu^2)}{y^3}\right]\nonumber\\ & & \label{eq:ope1b}
\eea
for $x>0$ and $x<0$ respectively. Now we multiply Eq. (\ref{eq:ope1a})
by $x^\beta$, integrate from $0$ to $1$
and let $\beta \rightarrow 0$, yielding
\be
\int_{0+}^1 \!\!\!\!dx h_L(x,\mu^2) = \int_{0+}^1 \!\!\!\!dx \left(h_1(x,\mu^2)
+h_L^3(x,\mu^2)\right)
+ \frac{m_q}{2M}\lim_{\beta \rightarrow 0} \beta \int_0^1 \!\!\!\!dx 
x^{\beta -1}
g_1(y,\mu^2), \label{eq:ope2a}
\ee
while multiplying Eq. (\ref{eq:ope1b}) by $|x|^\beta$ and integration from
$-1$ to $0$ yields
\be
\int_{-1}^{0-} \!\!\!\!\!dx h_L(x,\mu^2) = \int_{-1}^{0-}\!\!\!\!\! dx 
\left(h_1(x,\mu^2)
+h_L^3(x,\mu^2)\right)
- \frac{m_q}{2M}\lim_{\beta \rightarrow 0} \beta \int_{-1}^0 \!\!\!\!\!dx 
|x|^{\beta -1}
g_1(y,\mu^2). \label{eq:ope2b}
\ee 
Adding Eqs. (\ref{eq:ope2b}) and (\ref{eq:ope2a}) and
\begin{itemize}
\item assuming that there is no $\delta(x)$ in $h_1$, i.e. assuming that
\be
\int_{-1}^{0-}dx h_1(x)+ \int_{0+}^1dx h_1(x) = \int_{-1}^1 dx h_1(x)
\ee
\item assuming that there is no $\delta(x)$ in $h_L^3$ either and using 
that (from OPE), 
the lowest moment of $h_L^3$ vanishes identically i.e. assuming that
\be
\int_{-1}^{0-}dx h_L^3(x)+ \int_{0+}^1dx h_L^3(x) = \int_{-1}^1 dx h_L^3(x)=0
\ee
\item using that
\be\lim_{\beta \rightarrow 0} \beta \int_0^1 dx x^{\beta -1}
g_1(y,\mu^2) = g_1(0+,\mu^2)\ee
\end{itemize} 
one thus finds
\bea
\int_{0+}^1 dx \left[h_L^q(x,\mu^2) - h_L^{\bar{q}}(x,\mu^2)\right]
&\equiv& \label{eq:mismatch} 
\int_{-1}^{0-}dx h_L(x)+ \int_{0+}^1dx h_L(x)\\
&=& \int_{-1}^1 dx h_1(x,\mu^2) 
+\frac{m_q}{2M}\left[g_1(0+) - g_1(0-)\right] .\nonumber
\eea
Since the OPE also tells us that the first moments of $h_L$ and
$h_1$ ought to be the same (if $0$ is included in the integration), i.e.
$
\int_{-1}^1 dx h_L(x,\mu^2) = \int_{-1}^1 dx h_1(x,\mu^2)
$,
we thus conclude
\be
h_L(x,\mu^2) = h^{reg}_L(x,\mu^2) - 
\frac{m_q}{2M}\left[g_1(0+,\mu^2) - g_1(0-,\mu^2)\right]\delta(x) .
\label{eq:hreg}
\ee
A similar analysis applied to $g_T(x)$ yields \cite{mulders}
\be
g_T(x)=\int_x^1 \!\!\!\!dy \frac{g_1(y)}{y} + \frac{m}{M}\left[
\frac{h_1(x)}{x} - \int_x^1 dy \frac{h_1(y)}{y^2}\right]
+ \tilde{g}_T(x) - \int_x^1 \!\!\!\!dy \frac{\tilde{g}_T(x)}{y}
\ee
And hence under similar assumptions as above
\be
g_T(x,\mu^2) = g^{reg}_T(x,\mu^2) - 
\frac{m_q}{M}\left[h_1(0+,\mu^2) - h_1(0-,\mu^2)\right]\delta(x)
\label{eq:greg}
\ee
To summarize these results, although the OPE ensures that the
integrals of $h_L$ and $g_T$ are the same as those of $h_1$ and
$g_1$ respectively, this statement is strictly true only if the origin is
included in the integration. By analytic continuation of the moments,
we find that the behavior at the origin of $h_L$ and $g_T$ might be singular 
enough and the above statement about equality of the lowest moments 
seems to be violated if the origin is excluded from the integrals.
 
\section{A one-Loop Model for $h_L(x)$}
In order to illustrate the results from the previous section in a concrete
example, we consider $h_L(x)$ for a massive\footnote{In the previous section,
we found that the $\delta (x)$ term is explicitly proportional to the quark
mass quark!} quark in a one-loop model.

In such a model one finds for example
\bea
h_L(x)S^+ &\propto& \label{eq:1loop}\\
& &\!\!\!\!\!\!\!\!\!\!\!\!\!\!\!\!\bar{u}(P,S) 
\int \!\!\frac{d^2k_\perp dk^-
}{(2\pi)^3}
\gamma^\mu \frac{i}{\not \!\! k-m + i\varepsilon } 
\sigma^{+-} \gamma_5 
\frac{i}{\not \!\! k-m + i\varepsilon } \gamma^\nu u(P,S) D_{\mu \nu}(P-k)
\nonumber
\eea
where $k^+=xP^+$ and $D_{\mu \nu}(P-k)$ is the gluon propagator.
Here we have suppressed wave function renormalization terms
[$\propto \delta (x-1)$] because they are not relevant for the behaviour near
$x=0$.

In Eq. (\ref{eq:1loop}) the $\delta(x)$ terms can arise from terms with 
$k^-$ in numerator, because for those terms integrals
of the form \be \int dk^- \frac{k^-}{\left[k^2-m^2+i\varepsilon\right]^2}
\frac{1}{(p-k)^2+i\varepsilon}\label{eq:k-int}\ee
diverge linearly when $k^+=0$.
In order to see this, we rewrite
\be
k^- = P^- - \frac{\left({\vec P}_\perp-{\vec k}_\perp \right)^2}{
2(P^+-k^+)}
- \frac{\left(P-k \right)^2}{
2(P^+-k^+)}
\label{eq:algebra}
\ee
and note that the first two terms on the r.h.s. of Eq. (\ref{eq:algebra})
give ``regular''
expression when inserted into Eq. (\ref{eq:k-int}). However, the third
term on the r.h.s. of Eq. (\ref{eq:algebra}) cancels one of the
denominators in Eq. (\ref{eq:k-int}), yielding
\be
\int \frac{dk^-}{2\pi} \frac{1}{\left[k^2-m^2+i\varepsilon\right]^2}
= \frac{i}{2} \frac{\delta (k^+)}{{\vec k}_\perp^2 + m^2 } .
\ee
A detailed analysis \cite{me} shows that in $h_L$, 
the $\delta (k^+)$-terms survive, yielding a nonzero
$\delta (x)$ term that is proportional to $g_1(0+)$, \footnote{At one loop,
$g_1(x)=0$ for $x<0$ and thus $g_1(0-)=0$.}
which confirms Eq. (\ref{eq:hreg}).
A similar 1-loop analysis also shows that for $g_\perp$, 
the $\delta (k^+)$ term (at 1 loop) is multiplied
by $k^+$ $\Rightarrow$ no $\delta (x)$. This is consistent
with the result that $h_1^{1-loop}(0) = 0$ at one loop.

\section{Summary and Discussion}
At least in principle, one cannot exclude
subtractions in the dispersion relation between real \& imaginary
part of Compton amplitude. Whenever such a subtraction is present, this also
implies a $\delta (x)$-term in the corresponding parton distribution 
(defined as light-cone correlation).

A decomposition of $h_L$ and $g_T$ into twist-2 and twist-3 pieces
suggests that
$\lim_{\beta \rightarrow 0} \int_0^1 dx x^\beta g_\perp(x) 
\neq \int_0^1 dx g_\perp(x) 
$ and therefore 
$g_\perp (x)=g_\perp(x)^{reg} + c\delta (x) $.
A similar result is derived for $h_L$). 

The nonzero coefficient of $\delta(x)$ in $h_L(x)$
was confirmed by explicit one-loop calculations for
$h_L(x)$ as well as $g_1(0+)-g_1(0-)$.
At one loop, no $\delta(x)$ term was found in $g_\perp(x)$. 
This is consistent with the fact that
$h_1^{1-loop}(0)= 0$.

The prediction of these $\delta(x)$ terms from the moment relations 
[Eqs.(\ref{eq:hreg}),(\ref{eq:greg}], together with their one loop
verification, are the main result of this note.

Although the one-loop calculation yields $h_1(0)=0$, this result changes
at next to leading order (NLO)\cite{nlo}, i.e. $h_1^{NLO}(0)\neq 0$.
Therefore, using Eq. (\ref{eq:greg}), we expect that the
BC sum-rule \cite{bc} is also violated, but only at NLO.

In summary, both the $h_L$ sum rule as well as the $g_T$ are expected
to be violated if the point $x=0$ is not included
in the integrals.
For $h_L$ the violation appears already at leading order, while the
violation for $g_T$ does not appear until NLO. Both for $h_L$ and
$g_T$, the violation is proportional to $\frac{m_q}{M}$, i.e. for
$q=u,d$ the effect is expected to be very small. 
However, the contribution from strange quarks might be a significant.

Recently, it has also been suggested that $\delta(x)$ terms might be 
present in the twist-two distribution $g_1(x)$ \cite{bass}. 
Although there is not enough space here to discuss the results of Ref. 
\cite{bass} in detail, it should be emphasized that the general
features that have been discussed in this note also apply in that case.

One of the main differences between DIS and 
deeply virtual Compton Scattering (DVCS)
is that DIS measures only the
imaginary part of the forward Compton amplitude, while DVCS should
allow measurements of the full (real \& imaginary part\footnote{
The phase information can be obtained from interference with the 
Bethe-Heitler process.}) Compton amplitude ---
including non-forward matrix elements thereof.
The virtue of DVCS in this context is twofold. First of all, since also the
real part of the Compton amplitude is being measured, one could directly
test whether or not a subtraction appears in the dispersion relation.
Furthermore, any $\delta$-functions on forward parton distribution functions
get smeared out in non-forward distributions, i.e. the measurement of
the non-forward Compton amplitude acts as some kind of regularization.
Because of these features, studying near forward parton distributions using
DVCS may also be of help in clarifying the issues discussed in this paper.

\noindent {\bf Acknowledgements:}
It is a pleasure to thank X, Ji, Y. Koike and L. Mankie\-wicz
for helpful discussions. This work was supported in part by a grant from
DOE (DE-FG03-95ER40965), by a fellowship from JSPS  and in part by TJNAF.

\end{document}